# Systematic Investigation of the Intrinsic Channel Properties and Contact Resistance of Monolayer and Multilayer Graphene FET

Kosuke Nagashio*, Tomonori Nishimura, Koji Kita, and Akira Toriumi
Department of Materials Engineering, The University of Tokyo
7-3-1 Hongo, Bunkyo, Tokyo 113-8656, Japan
*E-mail: nagashio@maerial.t.u-tokyo.ac.jp

The intrinsic channel properties of monolayer and multilayer graphene were systematically investigated as a function of layer number by the exclusion of contact resistance using four-probe measurements. We show that the continuous change in normalized sheet resistivity from graphite to a bilayer graphene is governed by one unique property, i.e., the band overlap, which markedly increases from 1 meV for a bilayer graphene to 11 meV for eight layers and eventually reaches 40 meV for graphite. The monolayer graphene, however, showed a deviation in temperature dependence due to a peculiar linear dispersion. Additionally, contact resistivity was extracted for the case of typical Cr/Au electrodes. The observed high contact resistivity, which varies by three orders of magnitude (from ~$10^3$ to $10^6$ Ω μm), might significantly mask the outstanding performance of the monolayer graphene channel, suggesting its importance in future research.

## 1. Introduction

Graphene-based devices are promising candidates for future high-speed field-effect transistors (FETs), because a high carrier mobility of more than 10,000 $cm^2V^{-1}s^{-1}$ is reproducibly obtained by the mechanical exfoliation of bulk graphite.[1)] However, on-off current ratio ($I_{on}/I_{off}$) is small owing to the absence of bandgap. Several methodologies have been reported to improve this, such as quantum confinement using a small channel width for monolayer graphene[2,3)] and inversion symmetry braking using an external electric field for bilayer graphene[4,5)]. Therefore, the monolayer and bilayer graphenes have been the subjects of intensive focus, while only spot data are available for multilayer graphene. Since there has been no systematic investigation into the layer number dependence of transport properties from graphite to graphene, the scattered data on transport properties reported so far are not well organized. The monolayer graphene is a zero-gap semiconductor with a linear dispersion, while the multilayer graphene is a semimetal with a small band overlap between the conduction and valence bands.[6)] The defining characteristic of the multilayer graphene is band overlap, which is expected to decrease continuously from 40 meV for graphite.[7)] Therefore, it is important to understand comprehensively transport properties by estimating the band overlap of the multilayer graphene with a well-determined layer number, since the band overlap of graphite thin films thicker than 10 nm has been reported.[8)]

Another obstacle to a systematic understanding of graphene transport is that the transport properties obtained using a two-probe geometry, which include graphene/metal contact properties, are often discussed together with the intrinsic channel properties obtained using a four-probe geometry. The channel properties must be separated from the contact properties. Although the contact properties are crucially important, only a few experiments[9-12)] have addressed this matter because an ohmic contact is obtained without severe difficulty due to the lack of bandgap. Therefore, it is important to investigate the graphene/metal contact properties as well.

So far we have evaluated the layer number dependence of transport properties and reported room-temperature (RT) mobility as a function of layer number, which is accurately determined by combining optical contrast methods with Raman spectroscopy.[13)] In this study, temperature-dependent resistivity was measured for monolayer and multilayer graphene FETs with a typical four-probe geometry to estimate band overlaps. Moreover, contact resistance ($R_C$) for the typical metal contact of Cr/Au was also extracted systematically. Cr/Au electrodes were selected because they are the most commonly used. The objective of this study was to gain an understanding of the transport properties from graphite to graphene and the extrinsic graphene/metal contact properties.

## 2. Experimental Procedure

Graphite thin films were transferred onto 90 nm $SiO_2/p^+$-Si substrates by the micromechanical cleavage of Kish graphite. The number of layers was determined by optical contrast and Raman spectroscopy.[13-16)] Electron-beam lithography was utilized to pattern electrical contacts onto graphene. The contact metal [Cr (15 nm)/Au (20nm)] was thermally evaporated on resist-patterned graphene and subjected to lift-off in warm acetone. To remove the resist residual, graphene devices were annealed in a $H_2$-Ar mixture at 300 °C for 1 h.[17)] An optical micrograph of a typical graphene device with multi-probes is shown in Fig. 1(a). The electrical measurements were performed in a vacuum with a typical bias voltage of 10 mV while changing the temperature from 297 to 20 K.

## 3. Intrinsic Channel Properties

3.1 Layer number dependence of transport properties

Figure 2 shows (a) sheet resistivity ($\rho$) and (b) conductivity ($\sigma$) as a function of the gate voltage $V_g$ for the monolayer and multilayer graphenes. All data were obtained by four-probe measurement at RT. It should be noted that a back gate voltage of 30 V for 90 nm $SiO_2$ is almost equivalent to 100 V for 300 nm $SiO_2$. $\rho$ monotonically increases with a decrease in layer number, and then, for the monolayer graphene, the resistivity curve markedly changes and displays the smallest full-width at half-maximum. Similarly, the $\sigma$-$V_g$ curve for the monolayer graphene shown in Fig. 2(b) is sharper than the other curves. This fact is attributable to the much smaller density of states (DOS) for the monolayer graphene than for the multilayer graphene, due to the difference in dispersion relationship.

Figure 3(a) shows the FET mobilities estimated for the monolayer and multilayer graphenes. The mobility of graphene is generally estimated using the equations $\mu=1/en\rho$ and $n=C_{ox}(V_g-V_{Dirac})/e$, where $C_{ox}$ is the gate capacitance of $SiO_2$ and $V_{Dirac}$ is the gate voltage at the Dirac point, because the bulk carrier density determined by Hall measurement approximately agrees with the surface charge density [$C_{ox}(V_g-V_{Dirac})/e$] capacitively induced by the back gate. However, this is not the case for the multilayer graphene, especially for layer numbers of five or more, because the carrier density determined by Hall measurement is the sum of the surface charge density induced capacitively and the bulk carrier density (no electric field modulation). For the multilayer graphene, the substitution of only the surface charge density to $\mu=1/en\rho$ results in the overestimation of the mobility. In this study, the FET mobilities of the monolayer and multilayer graphenes were evaluated from the slope of the $I_d$-$V_g$ curve using

$$\mu = \frac{d_{ox}}{\varepsilon_{ox}} \frac{L}{W} \frac{1}{V_d} \frac{dI_d}{dV_g}, \qquad [1]$$

where $\varepsilon_{ox}$ and $d_{ox}$ are the permittivity and thickness of $SiO_2$, respectively. It is noted that, for the monolayer, the mobility determined using the equations $\mu=1/en\rho$ and $n=C_{ox}(V_g-V_{Dirac})/e$ reasonably agrees with that determined using eq.[1]. The mobilities of thin graphite films with 10-40 layers are roughly 8,000 $cm^2V^{-1}s^{-1}$. When the layer number was decreased, the mobility of the multilayer graphene with 2~8 layers decreased to ~3,000 $cm^2V^{-1}s^{-1}$. Finally, the highest mobility of 8200 $cm^2V^{-1}s^{-1}$ was achieved in the monolayer graphene at RT, although the data were widely scattered. The dotted line shows the trend in layer number dependence on mobility. This trend is similar to the typical values reported previously,[18,19] as shown by solid circles in Fig. 3(a).

For the mobility of the monolayer graphene, the transport is diffusive and not ballistic because the estimated mean free path of ~350 nm for the highest mobility of the monolayer graphene is much shorter than the device size of ~8 μm, as previously discussed.[13] It is hypothesized that the main factor for scattering is the remote Coulomb scattering caused by charged impurities that are expected to exist mainly at the graphene/$SiO_2$ interface.[20] This is because there is no Coulomb scattering center in the lattice of graphene due to the absence of impurity doping unlike in typical semiconductors, and the longitudinal acoustic (LA) phonon scattering in graphene is very weak.[21] Indeed, an extremely high mobility of more than 200,000 $cm^2V^{-1}s^{-1}$ has been reported for the suspended graphene, where charged impurities could be removed by Joule heating with current cleaning.[22]

The layer number dependence of the mobility is considered next. Figure 3(b) shows a schematic illustration for the monolayer, multilayer (2-8 L) graphenes, and the thin graphite films (~30 L). The charges induced by gate voltage are mainly located within a few layers because the screening length of the multilayer is reported as ~1 nm.[23-25] The induced carrier region is indicated by hatch marks. As shown by the table in Fig. 3(b), the major scattering factors appear to be remote Coulomb scattering at the top and bottom surfaces of the graphene and interlayer scattering. Generally, the effect of top-surface charged impurities should be less than that of bottom-surface charged impurities because it can be reduced by cleaning. In the multilayer graphene, all scattering factors affect the electric transport, resulting in a low mobility. On the other hand, in the monolayer graphene, the mobility is inherently high owing to the linear dispersion and lack of interlayer scattering. Electric transport in the monolayer graphene, however, should be sensitive to charged impurities due to the lack of additional layers to screen the charged impurities. It is suggested that the large mobility variation of the monolayer graphene is caused by variations in charged impurity density. If the charged impurity density can be reduced to a value below $4.8\times10^{11}$ $cm^{-2}$, the mobility will exceed 10,000 $cm^2V^{-1}s^{-1}$ at RT, as previously discussed.[13] Moreover, the mobilities of the thin graphite films are higher than those of the multilayer graphene because the distance between top-surface charged impurities and the carrier-induced region is long.[26] Although many other scattering factors can be expected to affect the transport, the trend of the layer number dependence of the mobility is explained by the main scattering factors shown in the table in Fig. 3(b).

The next topic of interest is the limitation of thickness for conductivity modulation. Thin graphite film can be divided into two regions as shown in Fig. 4(a); one region consists of a few layers near the interface where carriers can be induced by the electric field effect, and the other region is a bulk region where conductivity is constant. Therefore, total conductivity can be written as

$$\sigma_{total}(V_G) = \sigma_{FET}(V_G) + \sigma_{Bulk} \, .$$

In the case of $\sigma_{FET} > \sigma_{Bulk}$, the modulation of conductivity can be measured. In contrast, in the case of $\sigma_{FET} < \sigma_{Bulk}$, total conductivity is constant. Here, the minimum conductivity at the Dirac point ($\sigma_{min}$) is considered as a reference for $\sigma_{Bulk}$. Figure 4(b) shows $\sigma_{min}$ as a function of layer number at RT. It is clear that $\sigma_{min}$ increases with layer number. For a graphite film of ~70 nm thickness, the modulation of conductivity could not be detected experimentally. This thickness will be a limitation for the modulation, which is consistent with the previous experiment.[8)]

3.2 Temperature dependence of transport properties

Figure 5 shows the temperature dependence of $\rho$ for (a) the monolayer, (b) bilayer, and (c) four-layer graphenes. As the temperature decreased, the $\rho$ of the monolayer graphene decreased, while the $\rho$ values of the bilayer and four-layer graphenes markedly increased. To summarize the temperature dependence of $\rho$ for different layer numbers, Fig. 6 shows $\rho_{(T)}/\rho_{(297)}$ at the Dirac point as a function of temperature. The slope of $\rho_{(T)}/\rho_{(297)}$ continuously varies from negative to positive with increasing layer number from a bilayer to the bulk graphite. Particularly for two- to ten-layer graphenes, the change in slope is marked. For a graphite film with a thickness of 200 nm, $\rho_{(T)}/\rho_{(297)}$ was similar to that of the bulk graphite.[27)] From this continuous change in $\rho_{(T)}/\rho_{(297)}$ from graphite to the bilayer, it is reasonable to consider that the transport properties may be governed by one unique property, that is, the band overlap between the conduction and valence bands. However, only the monolayer graphene has almost no temperature dependence because the carrier density induced externally by the charged impurities is larger than the density of thermally excited carriers around the Dirac point due to the linear dispersion.[18)]

Figure 7 shows the temperature dependence of mobility for the monolayer and multilayer graphenes. The mobility exceeded 10,000 $cm^2V^{-1}s^{-1}$ at low temperatures for the monolayer graphene, while it was almost constant for the multilayer graphene (2-8 L). The mobility of the 12-layer graphene again increases with decreasing temperature.

Here, to understand the continuous change in $\rho_{(T)}/\rho_{(297)}$, it is assumed that the temperature dependence of FET mobility obtained in Fig. 7 is roughly the same as that of Hall mobility, especially for the multilayer graphene (2-8 L) because sheet resistivity depends on carrier density and Hall mobility ($\rho = 1/en\mu$). In this case, the continuous changes in $\rho_{(T)}/\rho_{(297)}$, $\mu_{(T)}/\mu_{(297)}$, and $n_{(T)}/n_{(297)}$ from graphite to a bilayer graphene can be schematically illustrated in Fig. 8. It should be noted that the monolayer graphene was excluded in this figure. The increase in $\rho_{(T)}/\rho_{(297)}$ with decreasing temperature for the multilayer graphene is due to the decrease[2] in $n_{(T)}/n_{(297)}$. Here, the carrier density of the bulk graphite can be written in a simple two-band model as

$$n = C_1 k_B T \ln\left[1 + \exp\left(\frac{\delta E_0}{2k_B T}\right)\right],$$

where $C_1$ is constant, $\delta E_0$ is the band overlap shown in the inset in Fig. 9(b), and $k_B$ is the Boltzmann constant.[7,28)] When a normalized carrier density is considered as $n_{(T)}/n_{(297)}$, there is only the single parameter of $\delta E_0$. Figure 9(a) shows $n_{(T)}/n_{(297)}$ obtained experimentally from Figs. 6 and 7. $\delta E_0$ values estimated by fitting the data [solid line in Fig. 9(a)] are shown in Fig. 9(b). It is evident that $\delta E_0$ markedly increases from 1 meV for the bilayer graphene to 11 meV for the eight-layer graphene. This marked change is well consistent with that of $\rho_{(T)}/\rho_{(297)}$. Now let us compare the present results with the data reported previously.[7, 8, 29,30)] Figure 9(b) shows that the present data is smoothly connected to those obtained by Novoselov et al.[7)], Zhang et al.[8)] and Moriki et al.[29)]. $\delta E_0$ seems to gradually reach 40 meV for the bulk graphite. However, Craciun et al.'s data[30)] is deviated from the present data. In their experiments, $SiO_2$ was deposited on the trilayer graphene for the top gate to control $\delta E_0$ by an electric field. Therefore, the damage to the trilayer graphene caused by the deposition may degrade the temperature dependence of $\rho$.

**4. Contact Properties for Cr/Au Electrodes**

In the previous section, the intrinsic channel properties of monolayer and multilayer graphene were extracted by excluding $R_C$ by the four-probe measurement. Here, in contrast, $R_C$ was extracted by

$$R_C = \frac{1}{2}\left(R_{2P} - R_{4P} \times \frac{l}{l'}\right),$$

where $l$ and $l'$ are the length between the source and the drain, and the length between two voltage probes, respectively, as shown in Fig. 1(b). Cr/Au electrodes was used for all the devices. Figure 10 shows the relationship between the contact resistivity ($\rho_C = R_C w$) and the ratio of two- and four-probe mobilities ($\mu_{4p}/\mu_{2p}$) for different layer numbers, where $w$ is the channel width. $\rho_C$ is varied widely from ~$10^3$ to $10^6$ Ω μm and seems independent of layer number. The minimum $R_C w$ in all the devices is ~$2\times10^3$ Ω μm. The value of $\rho_C$ = ~$1\times10^3$ Ω μm at $\mu_{4p}/\mu_{2p} = 1$ can be used as a reference value for good contact. The outstanding performance of the graphene channel might be significantly masked by the high contact resistivity, suggesting the importance of contact resistance in future research.

Figure 11 shows the temperature dependence of $\rho_C$ for the monolayer graphene. For $\rho_C$ > ~$10^4$ Ω μm, $\rho_C$ increases with decreasing temperature. The behavior of the thermal activation with an activation energy of ~0.2 eV was observed. Therefore, the temperature dependence of the sheet resistivity of the graphene channel (Fig. 6) should be measured using devices with a four-probe geometry, where the temperature dependence of the contact resistivity is excluded. Additionally, there was almost no temperature dependence for $\rho_C$ < ~$10^4$ Ω μm. Moreover, the variation in $\rho_C$ was suppressed to about one order of magnitude in this case.

As shown in Fig. 10, $\rho_C$ varies widely from device to device. However, the variations in contact properties in a single device with multiprobes have not yet been revealed. Therefore, the contact resistance was also estimated by the transfer length method (TLM) using the monolayer graphene with six electrodes, as shown in Fig. 12(a). All sets of two-probe resistances are plotted as a function of channel length in Fig. 12(b). The experimental results are reasonably fitted to a linear model, and $R_C$ is estimated as ~$5\times10^3$ Ω from the intercept indicating $2R_C$. Moreover, the $R_C$ estimated by TLM corresponds to the $R_C$ estimated by four-probe measurement using eq. [4]. These results suggest that the $R_C$ values of all metal electrodes are almost identical; that is, the variations in the contact properties of a single device are negligible. Moreover, it should be emphasized that $R_C$ has been properly extracted in the present study because $R_C$ does not depend on the methods used for its extraction.

Recently, electron-hole asymmetric resistivity, which is ascribed to charge transfer at the metal/graphene contact, has been discussed for Ti/Au electrodes.[9)] The work function difference between graphene and metal causes a strong charge transfer due to the small density of states for graphene, which results in p-doping (or n-doping) in graphene near the contact. Therefore, when a positive gate bias is applied, the p-n junction formed near the contact causes an additional resistance, while the increase in series resistance is negligible for a negative gate bias because of the absence of junction formation (p-p). This electron-hole asymmetric resistivity was first revealed by the comparison of the devices with invasive and external electrode geometries, as shown in Figs. 12(a) and 1(a), respectively, because the additional resistance caused by the p-n junction cannot be extracted by four-probe conductance measurement for the invasive case.[9)] Figure 13 shows $\rho_e/\rho_h$ as a function of carrier density for invasive and external Cr/Au electrodes. Compared with the reported value of $\rho_e/\rho_h$ =~2 for invasive Ti/Au electrodes,[9)] the deviation in $\rho_e/\rho_h$ from unity was within ±0.1, independent of probe geometry. Furthermore, Fig. 13 includes $\rho_e/\rho_h$ for the device "bcef" in Fig. 12 (a), where "b" indicates the source, "c" the drain, and "e" and "f" the two voltage probes. In this device, the number of p-n junctions is expected to be four, compared with two p-n junctions for the general case. $\rho_e/\rho_h$ was again close to unity, independent of the number of p-n junctions. These results suggest that the formation of the p-n junctions at the metal/graphene interface is negligible for Cr/Au electrodes because the difference in work function is negligible.[31)] Therefore, the Cr/Au electrodes are characterized by a high $\rho_C$ but without p-n junction formation.

## 5. Conclusions

We have systematically investigated the intrinsic channel properties of monolayer and multilayer graphenes by excluding $R_C$ using four-probe measurements. The continuous change in normalized sheet resistivity from graphite to a bilayer graphene is governed by one unique property, i.e., band overlap, which markedly increases from a bilayer to ~10 layers and then gradually reaches bulk graphite. The monolayer graphene, however, showed almost no temperature dependence because the carrier density at around $V_{Dirac}$ was controlled by the carrier density induced externally by charged impurities. On the other hand, in terms of the absolute value of mobility at RT, the layer number dependence is not very monotonic because both remote Coulomb scattering due to "extrinsic" charged impurities and "intrinsic" interlayer scattering affect the transport in multilayer graphene. Moreover, contact resistivity was extracted for the case of typical Cr/Au electrodes. The observed high contact resistivity, which varied by three orders of magnitude (from ~$10^3$ to $10^6$ Ω μm), might significantly mask the outstanding performance of the monolayer graphene channel, suggesting the importance of the metal/graphene contact to future research.

**Acknowledgements**

We would like to thank Professor S. Maruyama, University of Tokyo, for the use of the Raman spectroscope. The Kish graphite used in this study was kindly provided by Dr. E. Toya, Covalent Materials Co.

**References**

1) A. K. Geim and K. S. Novoselov: Nat. Mater. **6** (2007) 183.

2) M. Y. Han, B. Özyilmaz, Y. Zhang, and P. Kim: Phys. Rev. Lett. **98** (2007) 206805.

3) X. Li, X. Wang, L. Zhang, S. Lee, and H. Dai: Science **319** (2008) 1229.

4) J. B. Oostinga, H. B. Heersche, X. Liu, and A. F. Morpurgo: Nat. Mater. **7** (2008) 151.

5) Y. Zhang, T. T. Tan, C. Girit, Z. Hao, M. C. Martin, A. Zettle, M. F. Crommie, Y. R. Shen, and F. Wang: Nature **459** (2009) 820.

6) S. Latil and L. Henrard: Phys. Rev. Lett. **97** (2006) 036803.

7) K. S. Novoselov, A. K. Geim, S. V. Morozov, D. Jiang, Y. Zhang, S. V. Dubonos, I. V. Grigorieva, and A. A. Firsov: Science **306** (2004) 666.

8) Y. Zhang, J. P. Small, W. V. Pontius, and P. Kim: Appl. Phys. Lett. **86** (2005) 073104.

9) B. Huard, N. Stander, J. A. Sulpizio, and D. Goldhaber-Gordon: Phys. Rev. B **78** (2008) 121402(R).

10) P. Blake, R. Yang, S. V. Morozov, F. Schedin, L. A. Ponomarenko, A. A. Zhukov, R. R. Nair, I. V. Grigorieva, K. S. Novoselov, and A. K. Geim: Solid State Commun. **149** (2009) 1068.

11) S. Russo, M. F. Craciun, M. Yamamoto, A. F. Morpurgo, and S. Tarucha: arXiv:0901.0485v1.

12) E. J. H. Lee, K. Balasubramanian, R. T. Weitz, M. Burghard, and K. Kern: Nat. Nanotechnol. **3** (2008) 486.

13) K. Nagashio, T. Nishimura, K. Kita, and A. Toriumi: Appl. Phys. Express **2** (2009) 025003.

14) A. C. Ferrari, J. C. Meyer, V. Scardaci, C. Casiraghi, M. Lazzeri, F. Mauri, S. Piscanec, D. Jiang, K. S. Novoselov, S. Roth, and A. K. Geim: Phys. Rev. Lett. **97** (2006) 187401.

15) P. Blake, E. W. Hill, A. H. Castro Neto, K. S. Novoselov, D. Jiang, R. Yang, T. J. Booth, and A. K. Geim: Appl. Phys. Lett. **91** (2007) 063124.

16) Z. H. Ni, H. M. Wang, J. Kasim, H. M. Fan, T. Yu, Y. H. Wu, Y. P. Feng, and Z. X. Shen: Nano Lett. **7** (2007) 2758.

17) M. Ishigami, J. H. Chen, W. G. Cullen, M. S. Fuhrer, and E. D. Williams: Nano Lett. **7** (2007) 1643.

18) S. V. Morozov, K. S. Novoselov, M. I. Katsnelson, F. Schedin, D. C. Elias, J. A. Jaszczak, and A. K. Geim: Phys. Rev. Lett. **100** (2008) 016602.

19) S. V. Morozov, K. S. Novoselov, F. Schedin, D. Jiang, A. A. Firsov, and A. K. Geim: Phys. Rev. B **72** (2005) 201401(R).

20) S. Adam, E. H. Hwang, V. M. Galitski, and S. Das Sarma: Proc. Natl. Acad. Sci. U.S.A., **104** (2007) 18392.

21) J.-H. Chen, C. Jang, S. Xiao, M. Ishigami, and M. S. Fuhrer: Nat. Nanotechnol. **3** (2008) 206.

22) K. I. Bolotin, K. J. Sikes, J. Hone, H. L. Stromer, and P. Kim: Phys. Rev. Lett. **101** (2008) 096802.

23) P. B. Visscher and L. M. Falicov: Phys. Rev. B **3** (1971) 2541.

24) F. Guiena: Phys. Rev. B **75** (2007) 235433.

25) H. Miyazaki, S. Odaka, T. Sato, S. Tanaka, H. Goto, A. Kanda, K. Tsukagoshi, Y. Ootuka, and Y. Aoyagi: Appl. Phys. Express **1** (2008) 034007.

26) A. Kanda, H. Goto, and K. Tsukagoshi: Bull. Solid State Phys. Appl. **15** (2009) 114 [in Japanese].

27) D. E. Soule: Phys. Rev. **112** (1958) 698.

28) B. T. Kelly: Physics of Graphite, Applied Science, 1981, p. 285.

29) T. Moriki, A. Kanda, T. Sato, H. Miyazaki, S. Odaka, Y. Ootuka, Y. Aoyagi, and K. Tsukagoshi: Physica E **40** (2007) 241.

30) M. F. Craciun, S. Russo, M. Yamanoto, J. B. Oostinga, A. F. Morpurgo, and S. Tarucha: Nat. Nanotechnol. **4** (2009) 383.

31) G. Giovannetti, P. A. Khomyakov, G. Brocks, V. M. Karpan, J. van den Brink, and P. J. Kelly: Phys. Rev. Lett. **101** (2008) 026803.

**Figure captions**

**Figure 1** Optical micrograph (a) and schematic illustration of typical graphene FET with the multiprobes. The $SiO_2$ thickness used to increase the optical visibility of graphene is 90 nm. A $p^+$-Si substrate is used as a back gate.

**Figure 2** (a) Sheet resistivity and (b) conductivity as a function of gate voltage for graphene films at room temperature. "3L" indicates the trilayer graphene. It should be noted that a back gate voltage of 30 V for 90 nm $SiO_2$ is almost equivalent to 100 V for 300 nm $SiO_2$.

**Figure 3** (a) FET mobility variations for monolayer and multilayer graphenes with different layer numbers at room temperature. The dotted line shows the trend of the layer number dependence of mobility. Solid circles represent typical data reported previously.[18,19] (b) Schematic illustration and tables showing the main scattering factors for monolayer and multilayer graphenes, and thin graphite films. The induced carrier region is indicated by hatch marks.

**Figure 4** (a) Schematic illustration showing two regions, $\sigma_{FET}$ and $\sigma_{Bulk}$, in multilayer graphene. (b) $\sigma_{\mathrm{min}}$ as a function of layer number at room temperature.

**Figure 5** Temperature dependence of $\rho$ for (a) monolayer, (b) bilayer, and (c) four-layer graphenes.

**Figure 6** Sheet resistivity normalized to a 297 K value at the Dirac point as a function of temperature. The data for bulk graphite are also shown after ref. 27.

**Figure 7** Intrinsic FET mobility as a function of temperature for monolayer and multilayer graphenes.

**Figure 8** Schematic illustration of temperature dependence of normalized sheet resistivity, mobility, and carrier density from graphite to bilayer graphene. Noted that the monolayer graphene was excluded in this figure.

**Figure 9** (a) Normalized carrier densities obtained by experiment (solid circles) and fitting (solid line). (b) $\delta E_0$ values estimated by fitting the data in (a). The inset shows the schematic drawing of $\delta E_0$ for semimetal.

**Figure 10** Relationship between the contact resistivity and the ratio of two- and four-probe mobilities for different layer numbers.

**Figure 11** Temperature dependence of contact resistivity for monolayer graphene.

**Figure 12** (a) Optical micrograph of multiprobe graphene FET for the transfer length method. (b) Two-probe resistances as a function of channel length. The solid line shows the linear fit for the experimental data. Solid squares indicate $2R_C$ as estimated by four-probe measurement.

**Figure 13** $\rho_e/\rho_h$ as a function of carrier density for invasive and external Cu/Au electrode geometries. The solid triangle (four p-n junctions) is obtained for the device "bcef" in Fig. 12 (a). Experimental data for Ti/Au[9] is added in this figure for comparison.

# Figure 1

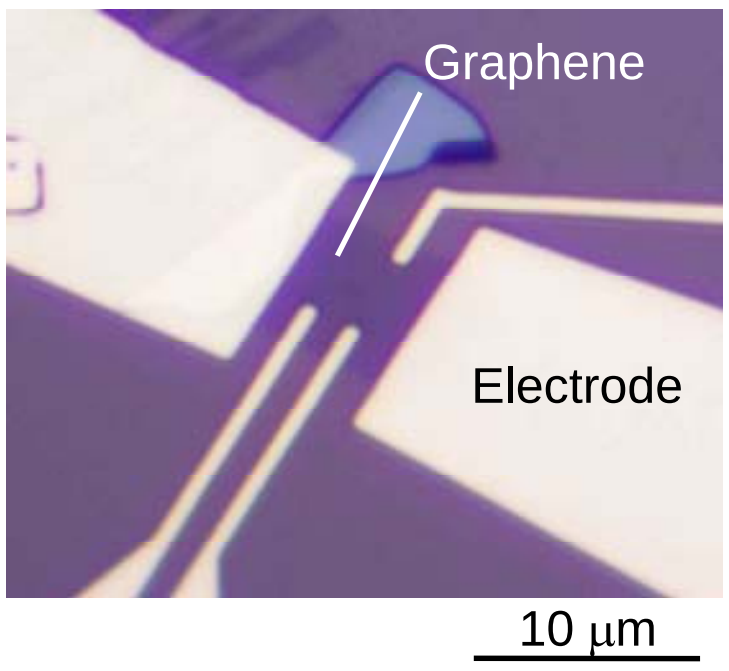

(a)

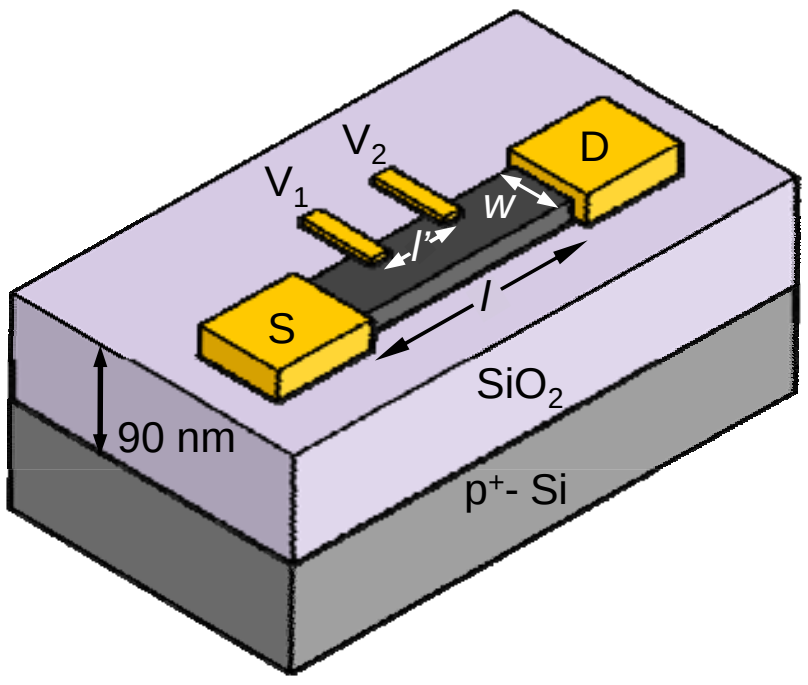

(b)

Nagashio et al.

# Figure 2

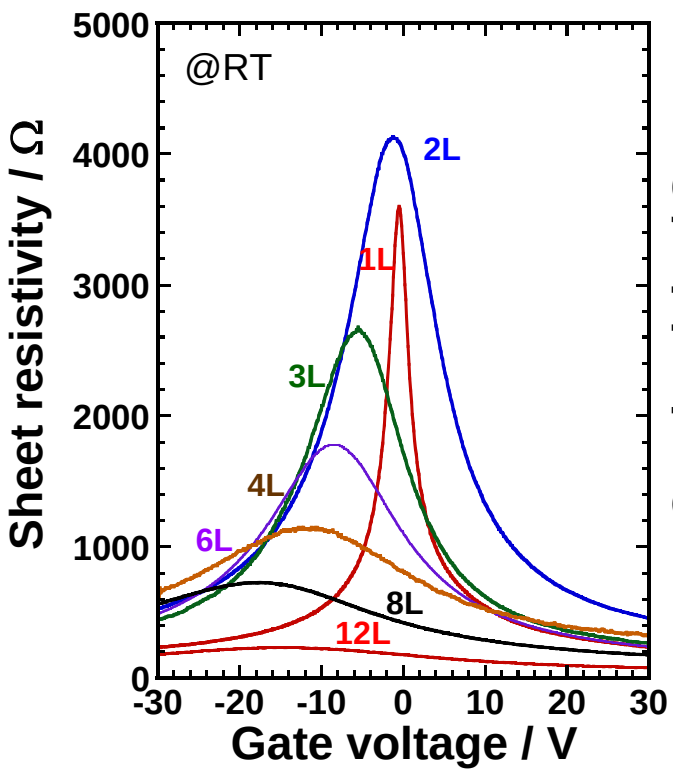

(a)

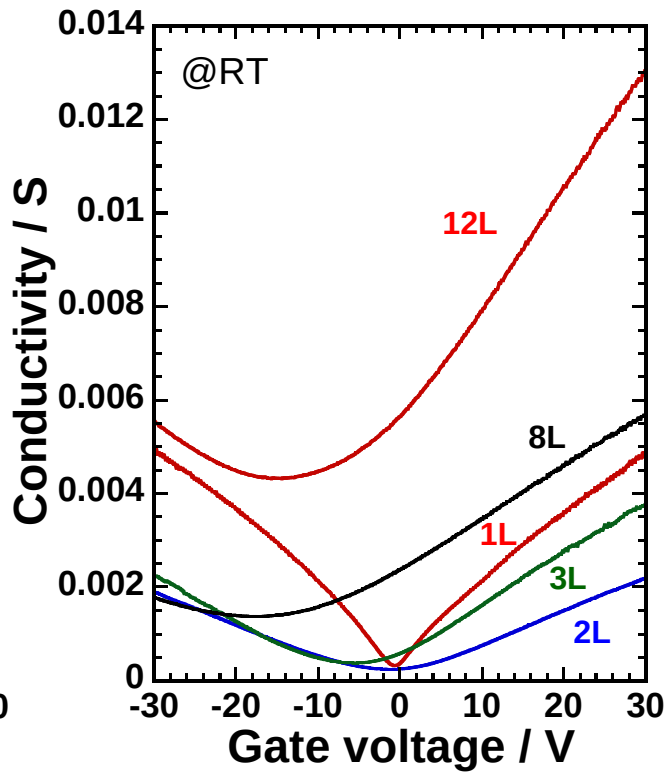

(b)

Nagashio et al.

# Figure 3

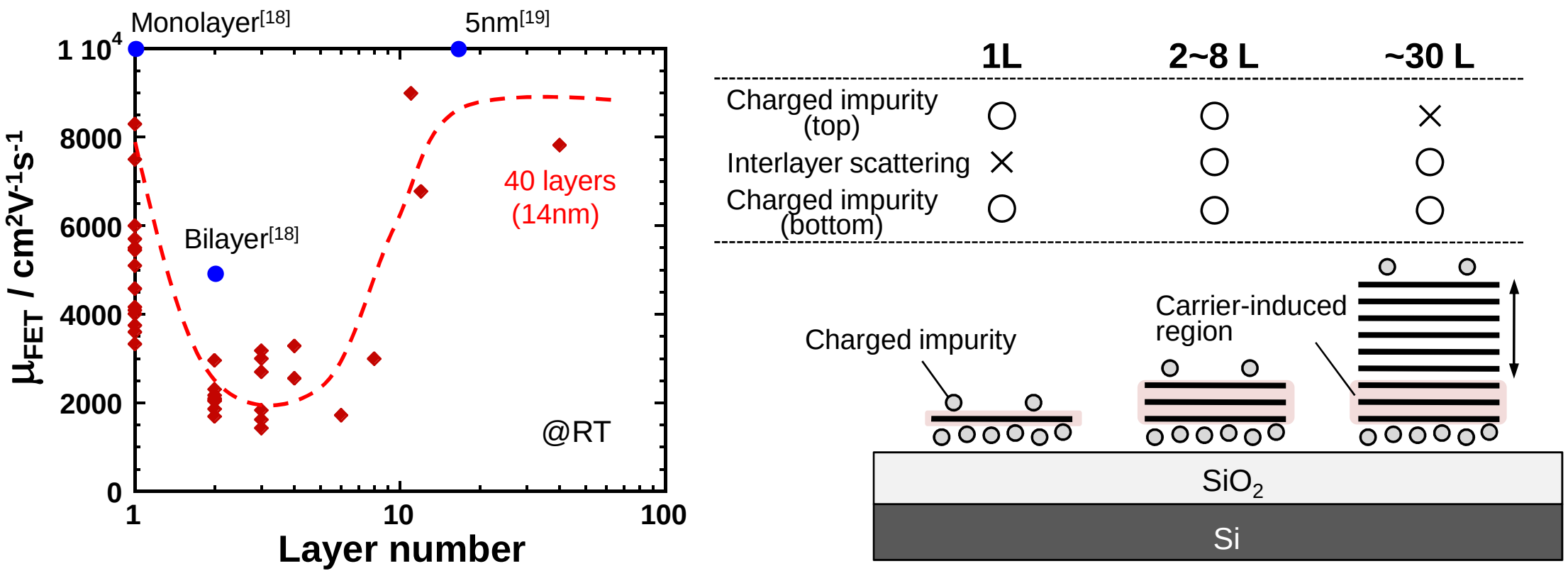

(a) (b)

Nagashio et al.

# Figure 4

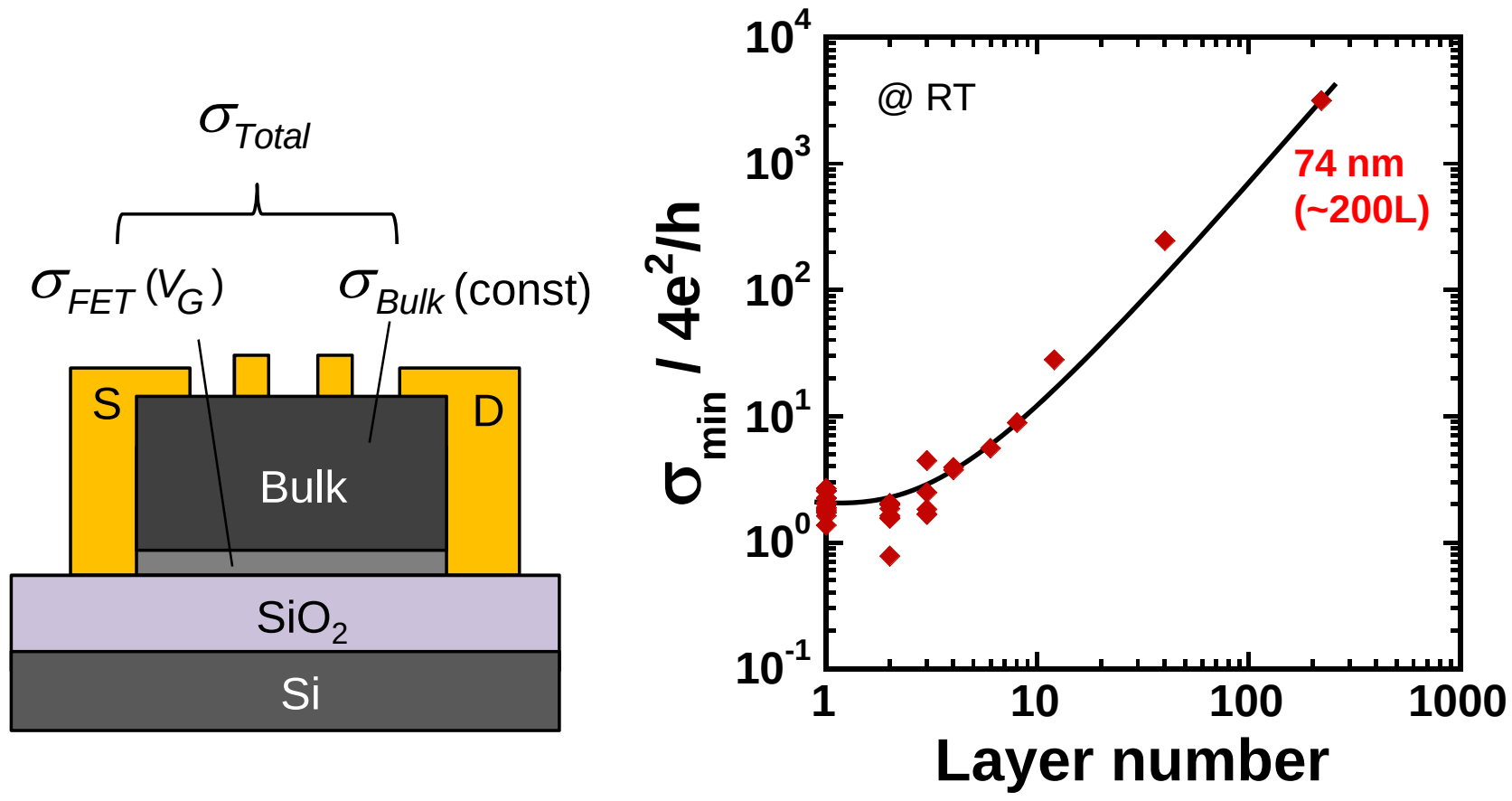

(a) (b)

Nagashio et al.

# Figure 5

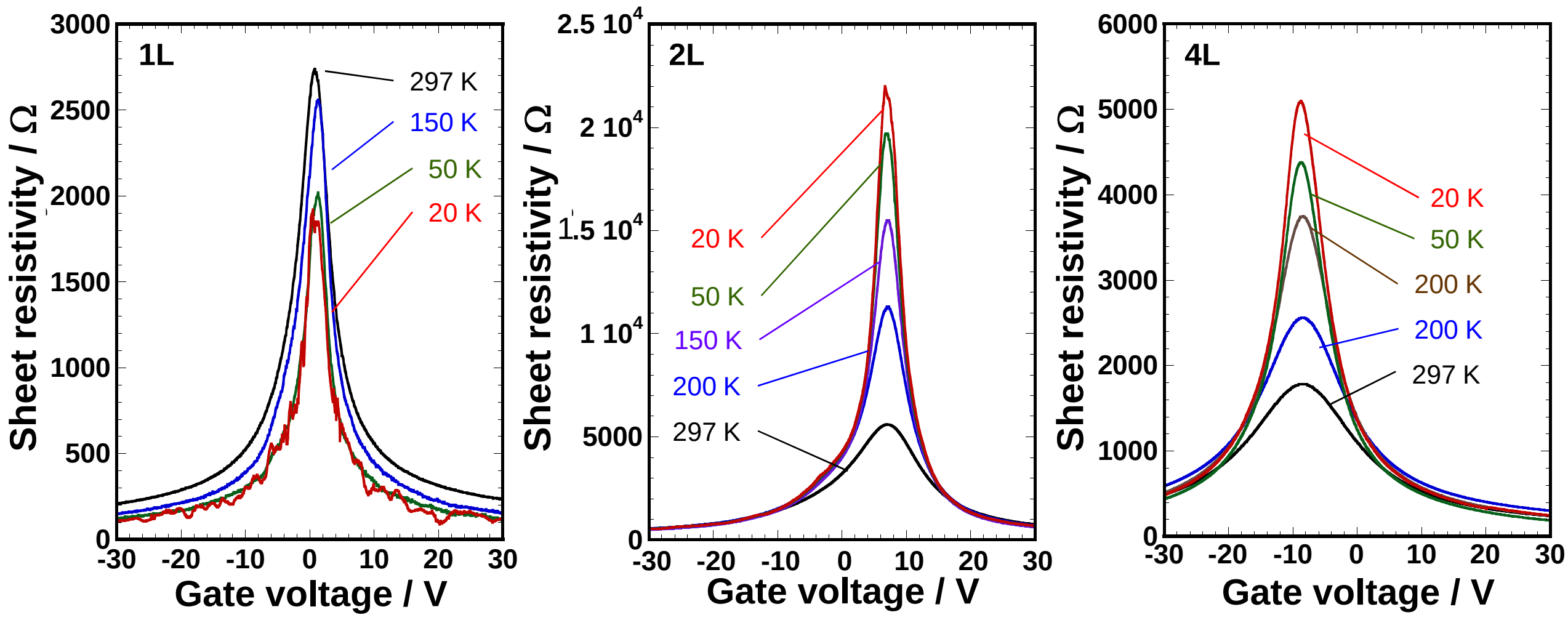

(a) (b) (c)

Nagashio et al.

# Figure 6

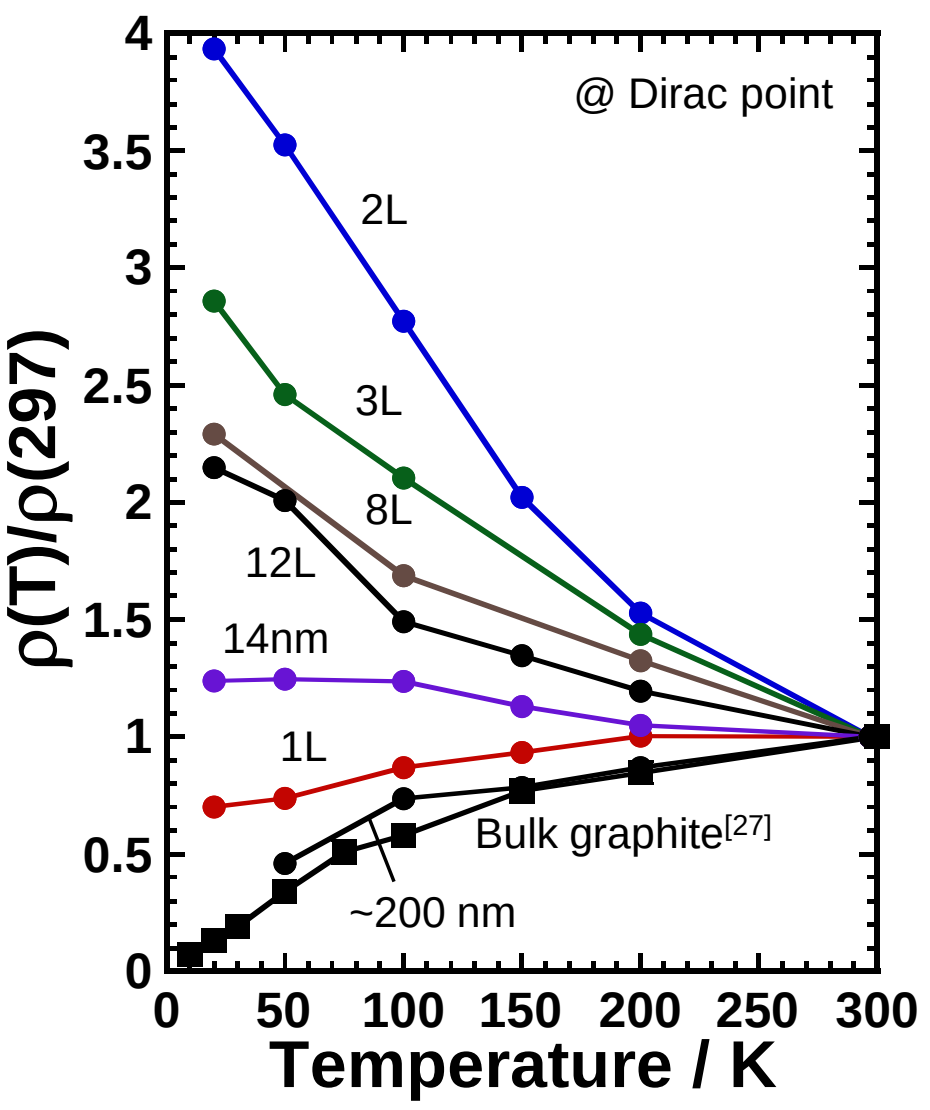

Nagashio et al.

# Figure 7

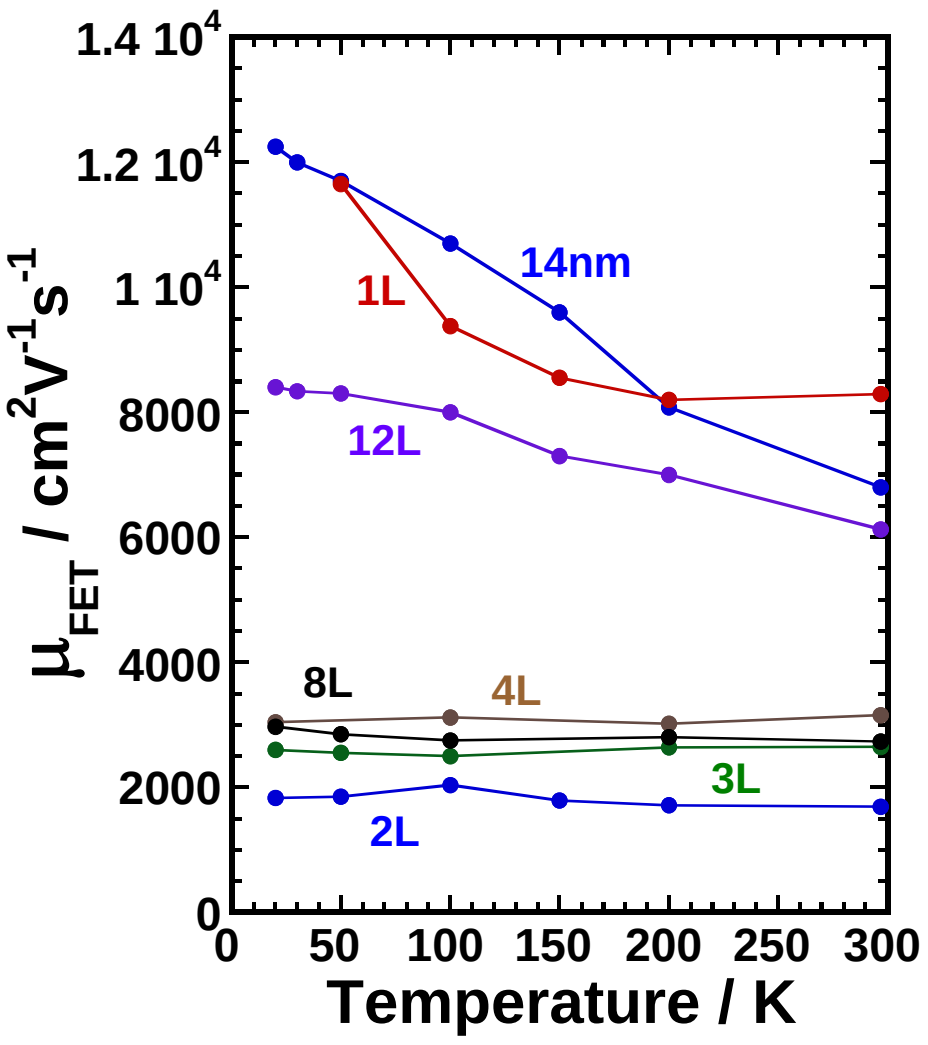

Nagashio et al.

# Figure 8

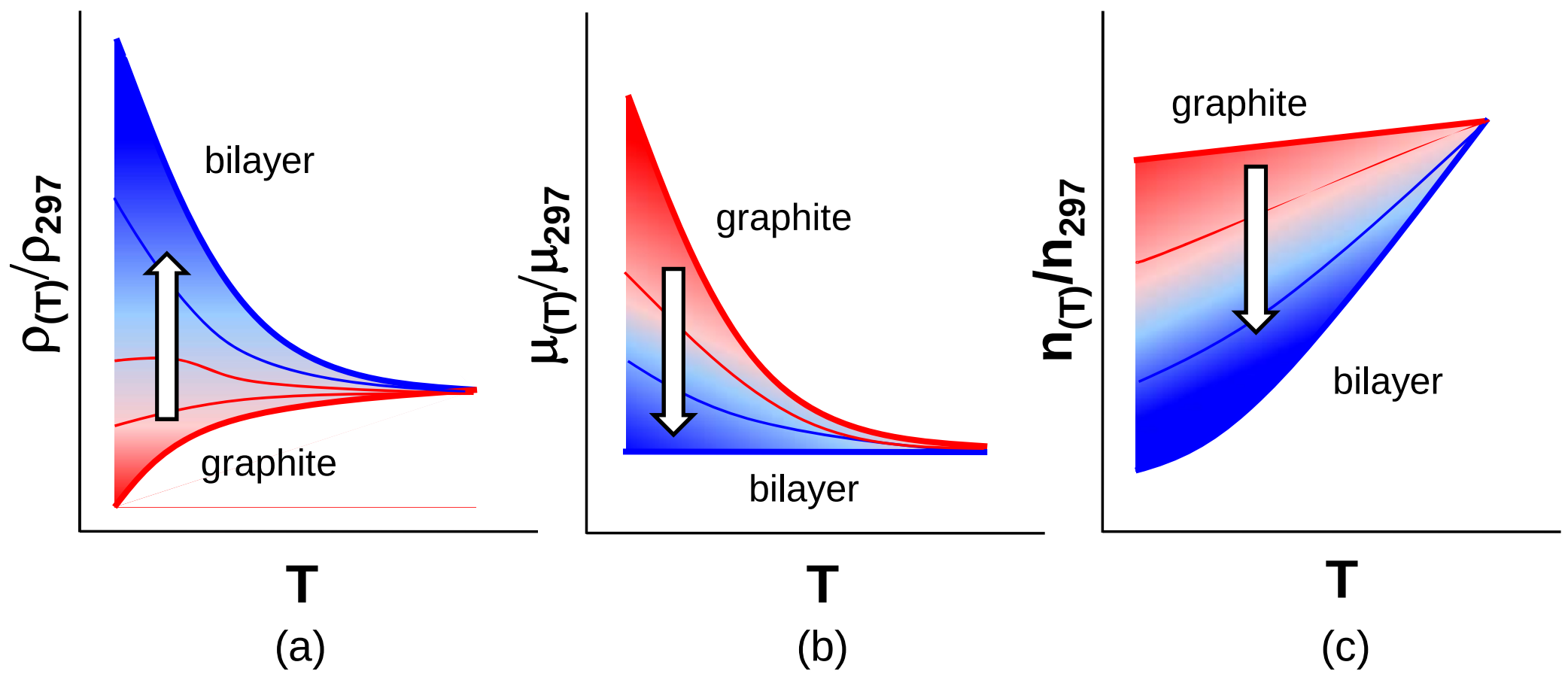

Nagashio et al.

## Figure 9

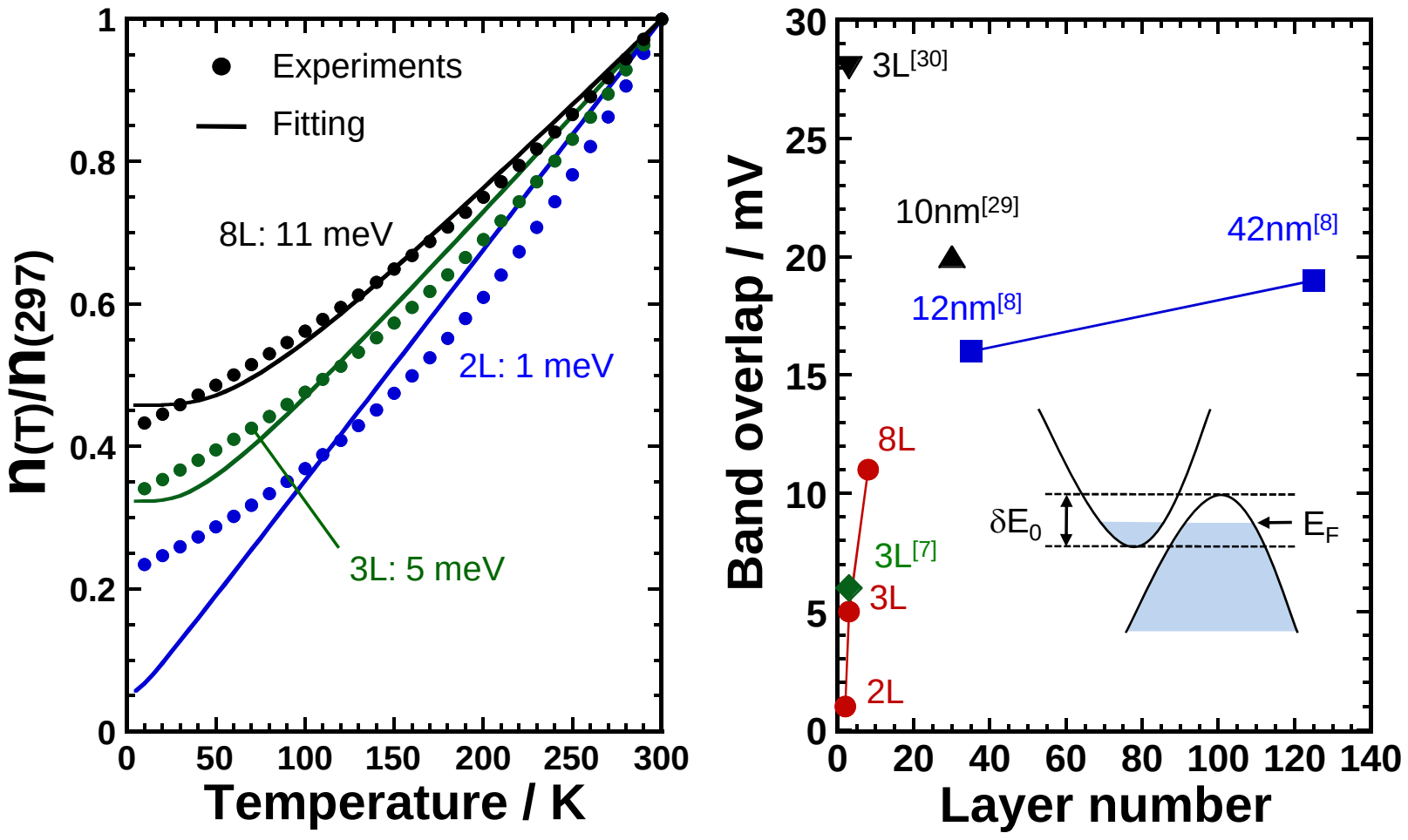

(a) (b)

Nagashio et al.

## Figure 10

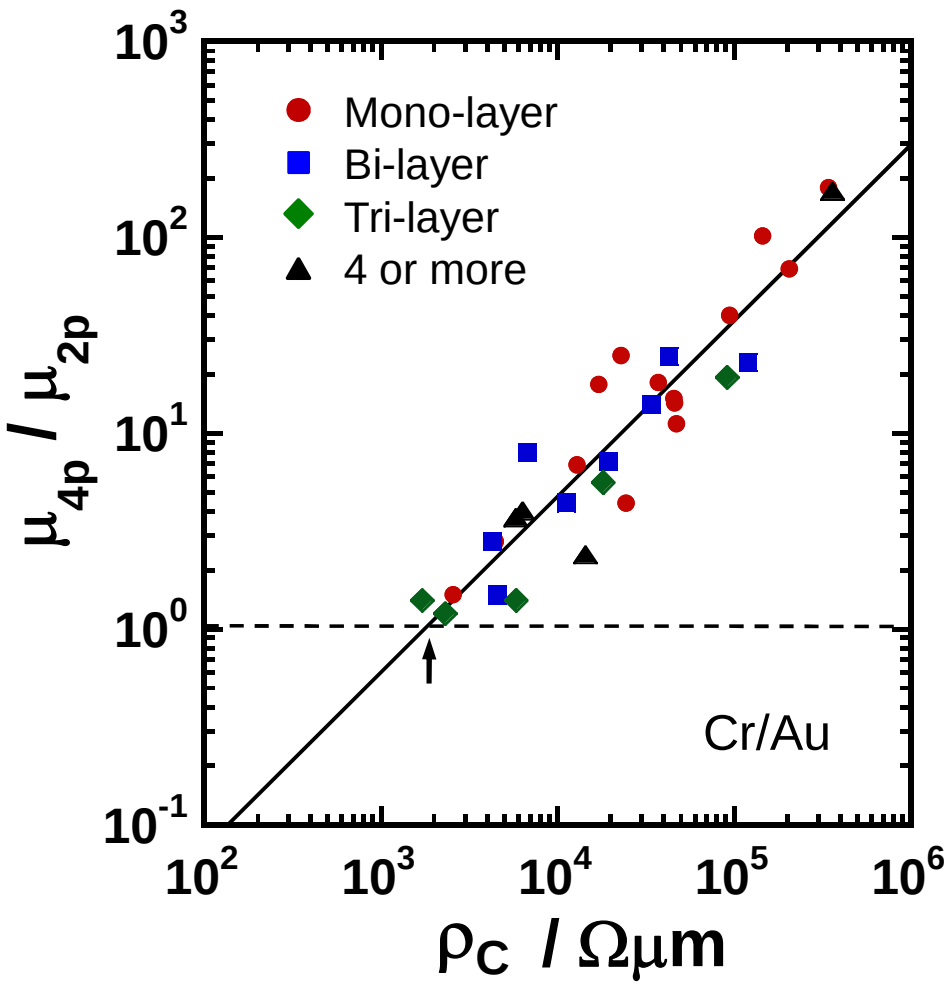

Nagashio et al.

# Figure 11

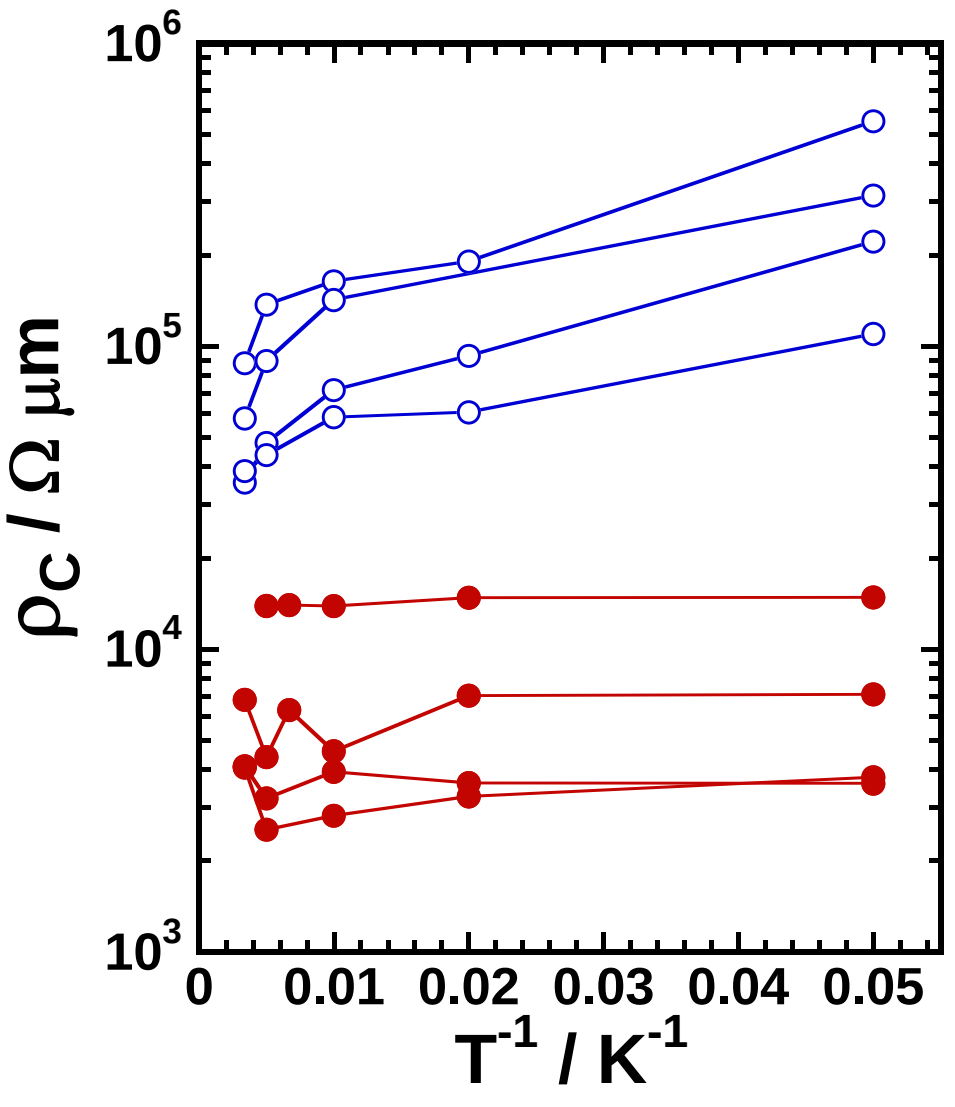

Nagashio et al.

# Figure 12

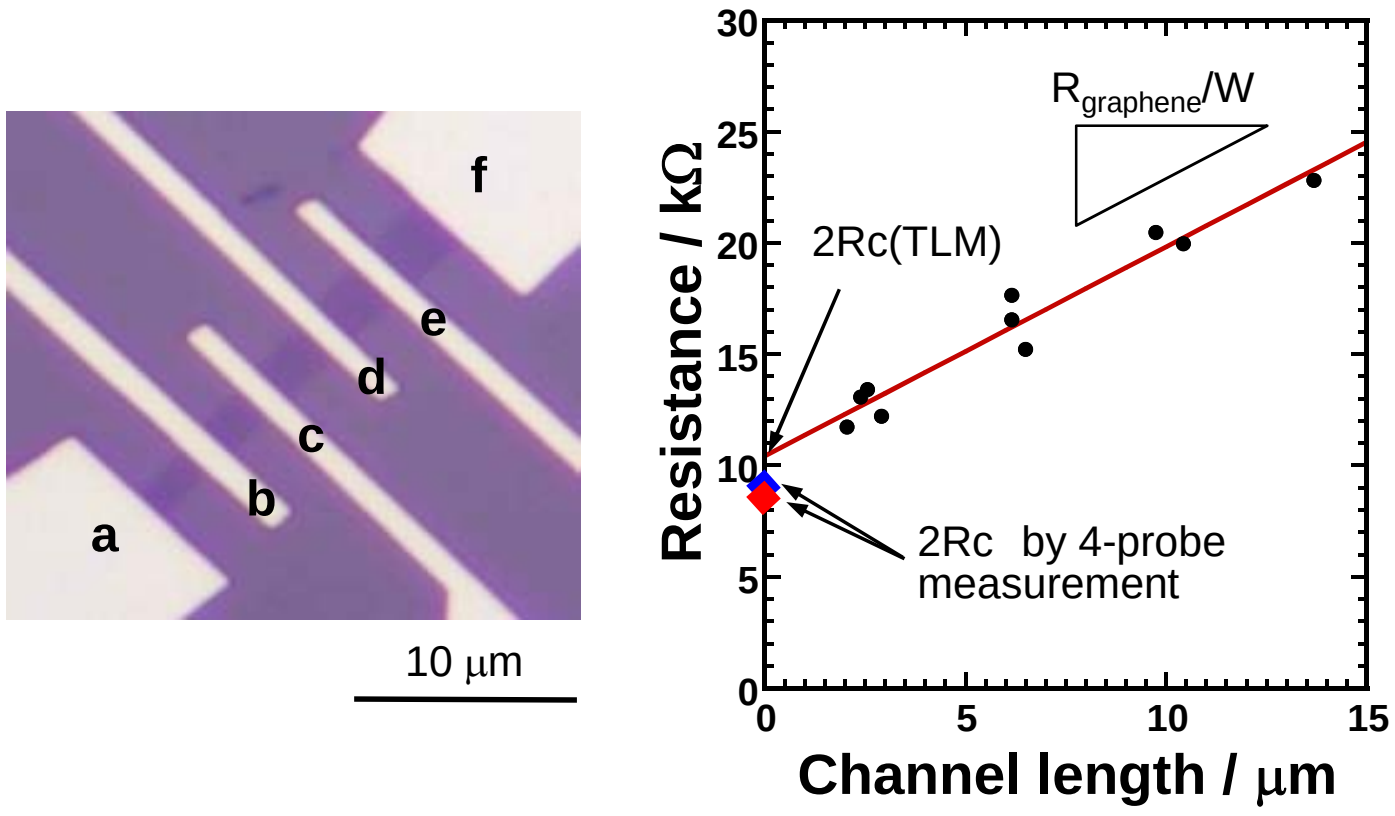

(a) (b)

Nagashio et al.

## Figure 13

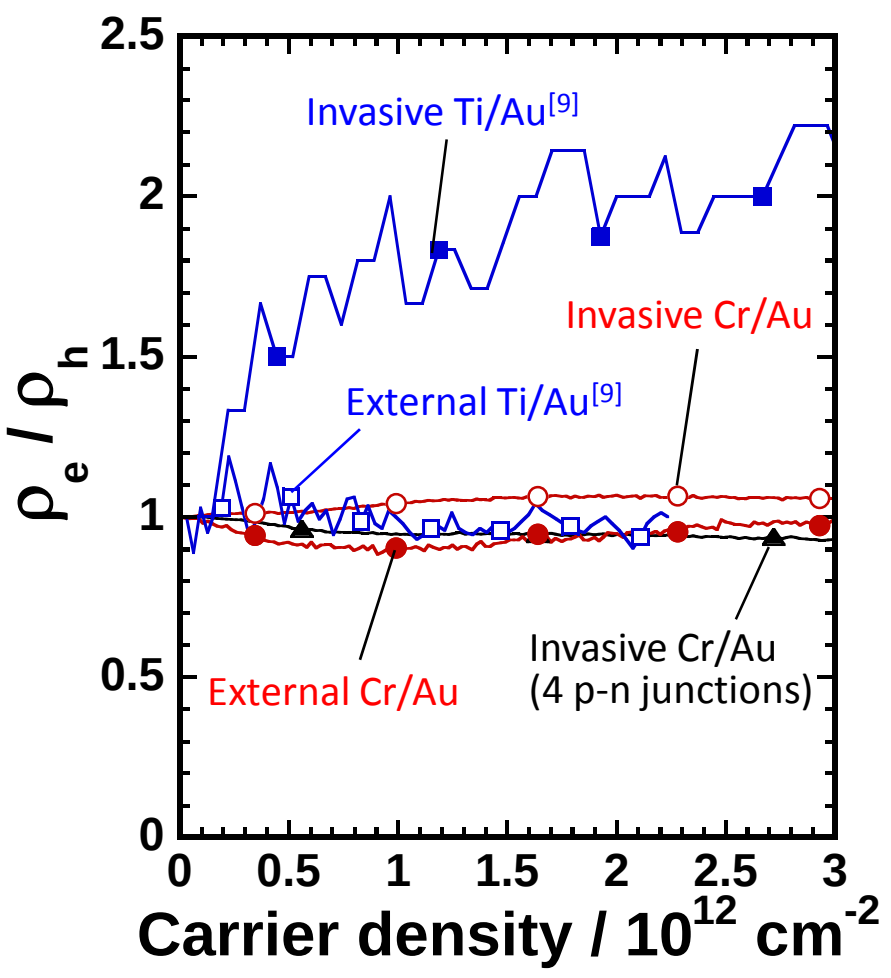

Nagashio et al.